\documentclass[12pt,english]{article}
\pdfoutput=1
\usepackage{graphicx}

\begin{document} 

\begin{center}

{\Large\bf The Pareto distribution as \\ \vspace{2mm} a disguised Gauss distribution} 

\vspace{3mm}

{\large\it Vladimir Pokrovskii}\footnote{ Vladimir Pokrovskii, vpok@comtv.ru} \\

\vspace{3mm}

{Moscow State University of Economics, Statistics and Informatics}  

\vspace{5mm}

\end{center}

\centerline{Abstract}

\vspace{5mm}

A simple heuristic model, including the  multiple exchanges between economic agents, is used to explain the mechanism of emerging  and maintenance of social  inequality in the market economy. The model allows calculating a density function of the population distribution over income. The function can be considered as a strongly deformed Gauss distribution function, whereas, at  large incomes, it coincides with the Pareto distribution. The external, in relation to the model under consideration, force is necessary to maintain the strong non-equilibrium in a stationary state, and this force  is the non-equivalence of elementary exchanges: the agent who already receives the higher income has the advantage: it provokes the rich to be getting more  rich and the poor to be  getting  pauper.

\vspace{5mm}
\vspace{5mm}

\vspace{5mm}


\newpage

 \section{Introduction}

At the end of the nineteenth century, sociologist and economist Wilfredo Pareto studied the distribution of the population of various countries over  income and the amount of wealth (Pareto, 1897, pages 299-345) and found that the distribution density function of the individuals $p(x)$ over  income $x$ in the region  of the large values is described  by a power function
\begin{equation}
p(x) = Ax^{-(1+\alpha)}.
\end{equation}
The power law (1) has been verified over and over again since then for different countries and at different times; the reliable estimates of the empirical values of  index $\alpha$ in the power distribution (1), according to Ribeiro (2020, p.84), range from 1.2 to 3.2.
 
The distribution (1), called the Pareto distribution, is not valid  in the region  of very small values of income, and the problem of finding the shape of the distribution function for those who do not belong to the rich class (the vast majority of people in modern societies), according to (Ribeiro, 2020, p.\,64), remained basically open. The presented  article refers to a simple heuristic model of a market economy, as an assembly of interacting economic agents, in order to calculate the distribution function with  the asymptotic Pareto distribution (1), and thereby to confirm  the mechanism of emergence and existence of the wealth  inequality. The author relies on a comprehensive  review of empirical results and theoretical approaches, which was recently presented by a Brazilian researcher (Ribeiro, 2020).

\section{Derivation of the distribution function}

\subsection{Heuristic model}

Following numerous studies (Ribeiro, 2020), we consider an  assembly  of   $N$ economic agents who jointly create products worth, say,~$Y$ per unit time. Each participant in production receives a share of the created  product -- the income of an individual $x$, which should be considered as a random variable, and, therefore, it is possible to introduce a distribution function of agents by income $p(x)$, such that the value of $p(x) \Delta x$ represents the number  of people with income between $x$ and $x + \Delta x$. Obviously, the following relations are fulfilled
$$
\int_0^\infty   p(x) d x = N, \quad  
\int_0^\infty  x  p(x) d x = Y
$$

Along with the distribution function of individuals over income, a distribution function of individuals over the their  contribution  into the  production can be introduced. In the simplest case, we can assume that the each agent inputs the equal amount of efforts  $Y/N$. In this case, the distribution function of individuals over  contribution to the production of value has the form of the  delta function  
$$
s(x) \sim \delta(x-Y/N) 
$$
It can be considered that the source with a power of $Y$ is located at the point $Y/N$ on the $x$ axis. It is assumed that the entire product is consumed  by the system of individuals. The assembly  of   $N$ economic agents ought to be considered as thermodynamic open system, the existence of which is supported by a flow of value (that can be reduced to a flow of energy) through the system.  

Since Pareto's time, many researchers have sought to understand whether the power distribution of wealth is a law of nature, or whether it can be reduced to more fundamental principles. The considered simple heuristic model, which takes into account the set of exchanges between economic agents according to market rules in the processes of production and distribution of value, allows us to calculate the distribution function of economic agents according to the income received and to establish the mechanism of occurrence and maintenance of economic inequality.

The market mechanism presupposes free exchange between various economic agents: the efforts of the workers  for money, money for products, products for other products, and so on. Under the  chaotic interactions, the income of an individual $x$ can be considered as a random variable, so that a change in the amount of income $a$  is distributed according to the  standard Gaussian form  
\begin{equation}
w\left(a, \langle a \rangle, \langle a^2 \rangle \right) = \left(\frac{1}{2\pi \langle a^2 \rangle} \right)^{\frac{1}{2}} \exp{\left[ -\frac{(a - {\langle a \rangle})^2}{\langle a^2 \rangle } \right] } 
\end{equation}
It is assumed also  that the transition probability function can depend on the variable $x$ through the  mean value ${\langle a\rangle}$ and standard deviation ${\langle a^2\rangle}$. The function $w(a, x)$ is defined as the probability of the agent's transition from the state $x$ to the state $x-a$. 

There is a method that allows us  to write a kinetic equation for the distribution function, based on the description of elementary transitions from one state to another (see, for example, Lifshitz and Pitaevsky, 1979, chapter~2). In this case, the change in the distribution function can be written as
\begin{equation}
\frac{\partial p(t, x)}{\partial t} = 
\int_{-\infty}^{+\infty} \left[ w(a, x+a) p(t, x+a) - w(a, x) p(t, x) \right] da \end{equation}
We assume that the change in the income $a$ is  much smaller than its current value, $a\ll x$; based on this, the first term in the integrand (3) can be replaced by the first terms of  expansion 
$$
w(a, x+a) p(t, x+a) \approx w(a, x) p(t, x) +  a \left. \frac{\partial (w p)}{\partial x}\right|_{a=0} + \, a^2 \left. \frac{\partial^2 (w p)}{\partial^2 x}\right|_{a=0}.
$$
The variables $a$ and $x$ are considered independent, and, because of this, the kinetic equation for the distribution function (3) can be written as
\begin{equation}
\frac{\partial p(t, x)}{\partial t} = \frac{\partial }{\partial x} \left[ \left(\langle a \rangle +  \frac{\partial \langle a^2 \rangle}{\partial x} \right) p + \langle a^2 \rangle \frac{\partial p}{\partial x} \right].
\end{equation}
Under made assumptions, the equation is valid in the region of positive values of $x$. If the function $w(x, a)$ is symmetric with respect to the variable $a$, $\langle a\rangle = 0$, and the probability variance does not depend on $x$, the first term on the right side of equation (4) disappears, and the kinetic equation reduces to a one-dimensional diffusion equation. To describe more complex situations, it should be assumed that the transition function may be asymmetric with respect to $a$, and the average value and variance  may depend on the variable $x$. 

Equation (4) defines a non-equilibrium distribution function and describes a balanced internal motion.   The square brackets on the right side of the equation (4) contain two terms: the diffusion leveling of the distribution of individuals by income is compensated by the flow of individuals.

\subsection{Steady-state distribution}

While  observing the results on inequality, Yakovenko and his collaborates  noticed, which is described in a short review (Ludwig and Yakovenko, 2021),  that the empirical distribution function can be represented as a composition of two functions: for small values of the variable, the distribution is described by the Boltzmann-Gibbs exponential function, while for the large values of the variable, the distribution corresponds to the Pareto power law. This gives them a reason to present the totality of individuals as quasi-independent communities of the poor (Ludwig and Yakovenko estimate it is more than 90\% in the USA) and the rich (about 4\% in the USA). Not to mention the fact that the two-component representation of distribution causes an acute sense of professional dissatisfaction, the scheme of Ludwig and Yakovenko omits the essence of the phenomenon: the mechanism of interchange between the poor and the rich, in the process of which wealth flows from the poor to the rich, and, one can think that  the representation of distribution by a single function gives a more adequate picture of the phenomenon.

In the steady-state case, the part of expression (4) enclosed in square brackets does not depend on $x$, and therefore we can introduce a constant $C$ and write the equation 
 \begin{equation}
\frac{d p}{d x} =\frac{1}{\langle a^2 \rangle}  
\left[C - \left(\langle a \rangle +  \frac{d \langle a^2 \rangle}{d x} \right) p \right]. 
\end{equation}
We are looking for a solution of this equation, which, for large values of $x$, ought to correspond to the Pareto distribution (1). The limit value of the function $p(\infty)$ is not known, but an expression for the derivative can be set as 
\begin{equation}
\left(\frac{d p}{d x} \right)_\infty = - \hat{p} (1+\alpha) x^{-(2+\alpha)}, \quad p = p(\infty) +  \hat{p} x^{-(1+\alpha)}.
\end{equation}

The  mean value ${\langle a \rangle}$ and the variance of ${\langle a^2 \rangle}$ in  equation (5), depend on income in the region of positive values of $x$. It can be expected that these quantities  are increasing functions of the values of $x$ (otherwise, exchanges of agents with high income may be inefficient), which are conveniently represented with power function. To satisfy  the asymptotes (6), we choose the dependencies
\begin{equation}
\langle a \rangle = \langle a \rangle_0 + r x^{1+\alpha}, \quad \langle a^2 \rangle = \langle a^2 \rangle_0 + k x^{2+\alpha}. 
\end{equation}
With these values of the indices, equation (5) has  the desirable  asymptotic (6),  and value of the constant in  equation (5)  could be found
\begin{equation}
C = \hat{p} \, [1 + r + (1+k)(1+\alpha)] , \quad   p(\infty) = 0. 
\end{equation}

Equation (5) with  relations (7) - (8) appears to be  an ordinary differential equation of the first order with the initial condition $p(0) = 0$, which we impose additionally. Without searching for an analytical expression of the solution, we turn to numerical methods that we use for the specific case of the distribution of individuals by income for Russia in 2018.

\subsection{An example from Russia}

Let us consider the income distribution density function for the population of Russia in 2018. In the Figure 1, the empty circles represent the values of the probability density of the distribution estimated  according to Rosstat data (Rosstat, 2023). One can notice that the dependence represents  typical income distributions (see, for example, Ribeiro, 2020, cc.\,93,\,98). In the same figure, a slight line shows  the Pareto power law (1) with index $\alpha = 1.65$, which reproduces empirical results  above the income $x > 50$.

Equation (5) is considered as an ordinary differential equation, which we solve numerically by the standard method. The simple consideration allows us to choose the parameters in relations  (7). For individuals with small incomes, the mean values are  close to zero, whereas the variance is not zero. Both quantities increase with income, but did not exceed the total amount of income, which determines the values of the parameters
\begin{equation}
\langle a \rangle_0 = 0; \quad r =  0.0000014. 
\end{equation}
The variance of the transition probability function is nonzero for small and large values of income and increases with increasing income according to the law (7) with the parameters 
\begin{equation}
\langle a^2 \rangle_0 = 0.02; \quad  k = 0.0000005.  
\end{equation}

The value of $C$ is determined by the relation  (8), while  the parameter $\hat{p}$ can take an arbitrary sufficiently large value; calculations are carried out at a value of $C = 36.6$.

We are looking for a solution of  equations (5) - (8) with the values of the parameters (9) and (10) that are specific values for the example.   The result of calculating the distribution function is shown in the Figure 1 in comparison 
with the empirical values and the Pareto distribution. The graph of density 
\linebreak
\hspace*{4mm}
\centerline{\includegraphics[scale=0.5, bb=0 0 650 350]{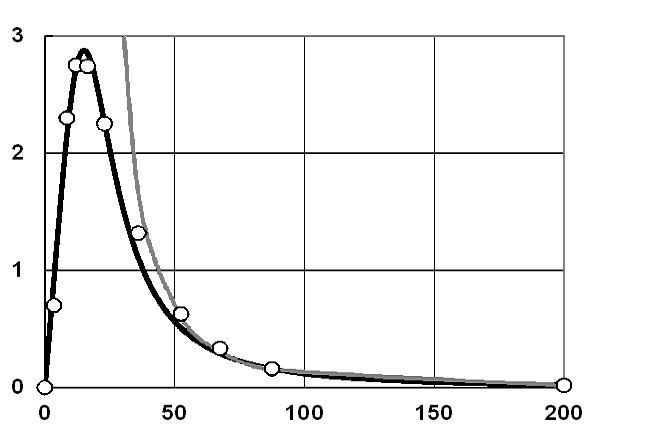}} 

\vspace{5mm}

\centerline{\bf \ Figure 1 Distribution density function}
\vspace{2mm}
{\it\small\noindent The empty circles show empirical values of the density of distribution of the Russian population by income in 2018, calculated  according to available data (Rosstat, 2023). The slight line represents the Pareto distribution (1) at $\alpha = 1.65$. The solid line shows  the solution of equation (5) at the values of the parameters (9) and (10). The distribution density function is normalized by 100 units. Income is  shown  in thousands of rubles.}

\vspace*{3mm} 
\hrule 
\vspace*{6mm} 
\noindent
function of the distribution represents a strongly deformed Gaussian distribution function, and this result convinces us that there is no need to introduce a quasi-equilibrium two-component scheme either on formal or substantive grounds.

\section{The anatomy of inequality}

The share of the created product that is received by an individual  does not correspond, generally speaking, to the contribution of the individual to the production of value. Now,  returning  to the consideration of an assembly of individuals, each of whom, according to our simplest assumption, produces the  value of $Y/N$ units, so that distribution of individuals over  participation in production \ is represented \ by delta-function (see Section 2.1). \  Each \  individual receives its share of the created product, however,  the distribution of individuals over  income could  be different depending on the circumstances 
\linebreak


\centerline{\includegraphics[scale=0.28, bb=0 0 1050 550]{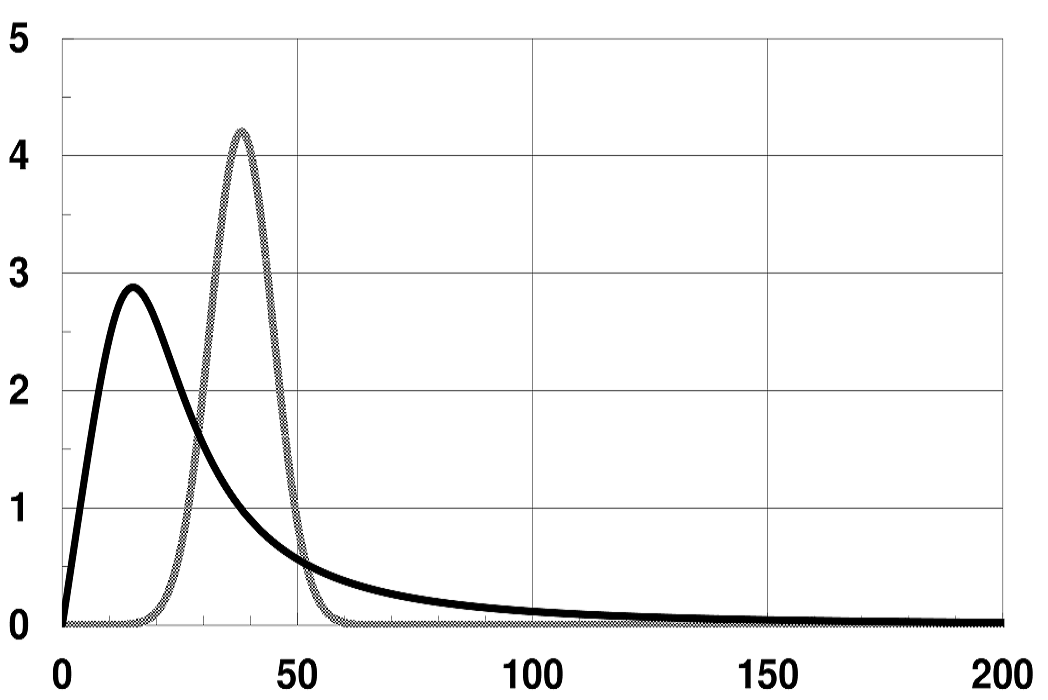}}

\vspace{5mm}

\centerline{\bf Figure  2 \ Income distribution functions}
\vspace{2mm}
{\it\small \noindent The slight  curve represents the natural distribution of individuals over income, calculated for  the average value of contribution  by an individual $Y/N=38.1$ and variance $\langle (x - Y/N)^2 \rangle = 90$. The solid curve represents the real distribution of the Russian population by income in 2018, reproduced from Figure 1. The distribution functions are normalized by 100 units. Income is presented in thousands of rubles.}
\vspace*{3mm} 
\hrule 
\vspace*{4mm} 
\noindent
of the allocation. In a simple case, when there are only random interactions of individuals and other chaotic uncontrolled influences, the probability function of the distribution of individuals over  income takes the form of the standard Gaussian distribution about  the mean  value of income $Y/N$
\begin{equation}
p\left(x \right) = \left(\frac{1}{2\pi \langle (\Delta x)^2 \rangle} \right)^{\frac{1}{2}} \exp{\left[ -\frac{(\Delta x)^2}{\langle (\Delta x)^2 \rangle } \right] }.  
\end{equation}
Here it is denoted $\Delta x = x - Y/N$. The graph of the  distribution is shown 
in Figure 2 by the slight line.

However, the situation in reality appears to be more complicated. The approximation  for the distribution function of individuals over by contribution to the production of value by delta-function seems reasonable, but the real distribution function of individuals over income differs significantly from the Gaussian distribution and has the form (1) in the region of big incomes.  Figure 2 shows a graph of the real distribution function of individuals by income, reproduced from Figure 1.

The real income distribution looks like a deformed Gaussian distribution. Both one and the other distributions shown in Figure 2 are stationary non-equilibrium distributions. Turning to the situation in Russia, we note that, according to the assumption, each individual, regardless of his real income, produces $Y/N = 38.1$ units of value. It can be easily estimated that about 70\% of individuals have a real income less than the average value of $Y/N =38.1$; in total they receive about 35\% of the entire product. The remaining part of individuals, respectively, receives an income exceeding the mean  value, which means that there is a mechanism that transfer some amount  of the value from individuals with lower income to individuals with higher income. Such a flow of value cannot arise naturally, without some influences that are  external in relation to the model under consideration. There is an embodied mechanism, which  supports the  trends towards the non-equivalence of elementary exchanges: in  each act, the agent who already receives the higher income has an advantage. Formally, the non-equivalence of the elementary exchanges is related to the asymmetry of the  transition function between states (2) with a non-zero mean transition value.

\section{Conclusion}

Thus, it can be argued that the cause of property inequality is that there are such rules for making exchanges (a market mechanism), in which the rich get richer and the poor become poor. The article describes the mechanism of the emergence and maintenance of economic inequality on the basis of a very simple model, but does not assess the existence of inequality and does not find out whether this contributes to a favorable life of society or not. This is a different  problem that  could  be considered on the basis of the clarified patterns of exchange, which can be used as the basis for mathematical models that determine the impact of inequality on economic activity.

\newpage

\section*{References}

\begin{description}

{\small 
\item Pareto V. (1897) {\it Cours d'economique politique. The first edition}.  Macmillan, London. Also: Pareto V. {\it Cours d'Economie Politique: Nouvelle edition \linebreak 
par G.-H. Bousquet et G. Busino}, Librairie Droz, Geneva, 1964. 

\item  Ribeiro M.B. (2020) Income distribution dynamics of economic systems:
An econo-physical approach. Cambridge,  UK:  Cambridge University Press.

\item Lifshitz E.M. and Pitaevskii L.P. (1981), Physical Kinetics. Oxford: Pergamon,.

\item Ludwig D. and Yakovenko V.M. (2021) Physics-inspired analysis of the two-class income distribution in the USA in 1983-2018. Phil. Trans. R. Soc. A 380, 20210162. (doi:10.1098/ rsta.2021.0162) 23.  arXiv:2110.03140v1 [physics.soc-ph] 7 Oct 2021

\item Rosstat (2023)   Population by average per capita money income. Table: urov\_31.xlsx (updated 26.12.2019). Available at: https://www.gks.ru/ (accessed: 16 February 2023).
}

 \end{description}

\end{document}